# The viscoelastic properties of chromatin and the nucleoplasm revealed by scale-dependent protein mobility

**Fabian Erdel, Michael Baum & Karsten Rippe**

Deutsches Krebsforschungszentrum (DKFZ) and BioQuant, Research Group Genome Organization & Function, Im Neuenheimer Feld 280, 69120 Heidelberg, Germany

E-mail: F.Erdel@dkfz.de or Karsten.Rippe@dkfz.de

**Abstract**. The eukaryotic cell nucleus harbors the DNA genome that is organized in a dynamic chromatin network and embedded in a viscous crowded fluid. This environment directly affects enzymatic reactions and target search processes that access the DNA sequence information. However, its physical properties as a reaction medium are poorly understood. Here, we exploit mobility measurements of differently sized inert green fluorescent tracer proteins to characterize the viscoelastic properties of the nuclear interior of a living human cell. We find that it resembles a viscous fluid on small and large scales, but appears viscoelastic on intermediate scales that change with protein size. Our results are consistent with simulations of diffusion through polymers and suggest that chromatin forms a random obstacle network rather than a self-similar structure with fixed fractal dimension. By calculating how long molecules remember their previous position in dependence on their size, we evaluate how the nuclear environment affects search processes of chromatin targets.

# 1. Introduction

The interior of the cell nucleus represents a complex medium that on the one hand has properties of a liquid but on the other hand contains a solid chromatin scaffold and large macromolecular aggregates that can dynamically change their three-dimensional conformation (Wachsmuth *et al.*, 2008). Chromatin itself comprises to a large extent highly variable and often irregular structures and thus has been referred to as a 'polymer melt', emphasizing its complex material properties (Maeshima *et al.*, 2014). Despite the lack of membrane-confined subcompartments, the nuclear interior constrains transport of biological macromolecules like proteins and RNA molecules that are central to the readout and replication of the genetic information (Hofling and Franosch, 2013). The properties and microscopic origins of the resulting anomalous diffusion processes have received considerable interest (Saxton, 1994, 1996; Condamin *et al.*, 2008) since they have a large impact on enzymatic reactions that occur in the nucleus (Benichou *et al.*, 2010; Guigas and Weiss, 2008; Bancaud *et al.*, 2009; Lieberman-Aiden *et al.*, 2009; Woringer *et al.*, 2014).

One important factor that needs to be considered in this context is the topology of chromatin as the dominating structural scaffold in the nucleus (Bancaud *et al.*, 2009; Bancaud *et al.*, 2012; McNally and Mazza, 2010; Lieberman-Aiden *et al.*, 2009; Mirny, 2011; Dekker *et al.*, 2013; Baum *et al.*, 2014). Many models describe chromatin as a fractal, and its apparent fractal dimension has been assessed in experiments and theory (e.g. (Lieberman-Aiden *et al.*, 2009; Mirny, 2011; Tajbakhsh *et al.*, 2011)). In this type of studies the conformation of chromatin is described by simple scaling laws according to distinct polymer models and thus has fractal properties (Flory, 1969; Witten, 1998; Rudnick and Gaspari, 1987). These considerations have prompted the hypothesis that also the nuclear space might be fractal (Bancaud *et al.*, 2009; Bancaud *et al.*, 2012; McNally and Mazza, 2010). Such an organization would impose specific constraints on transport processes that occur within the cell nucleus (Benichou *et al.*, 2010; Guigas and Weiss, 2008). However, both the conclusion that chromatin can be described by a simple polymer conformation of fractal nature and the assumption that this is related to a fractal environment for soluble nuclear factors have been challenged (Nicodemi and Pombo, 2014; Baum *et al.*, 2014). Additionally, the theoretical expectation would be that the complement of a fractal is not necessarily fractal (Crawford and Matsui, 1996; Mandelbrot, 1982).

One approach to deduce the organization of the nuclear interior is it to study particle mobility and accessibility of labeled components by fluorescence correlation spectroscopy (FCS), fluorescence recovery after photobleaching (FRAP) or single particle tracking (SPT) experiments (Bancaud *et al.*, 2009; Dross *et al.*, 2009; Pack *et al.*, 2006; Moeendarbary *et al.*, 2013; Grunwald *et al.*, 2008). Both experimental and theoretical studies on transport processes inside polymers and cells demonstrate that the intracellular diffusion behavior is anomalous (Fritsch and Langowski, 2010; Hofling and Franosch,

2013; Wachsmuth *et al*., 2000; Weiss *et al*., 2004). However, the limited range of length/time scales of previous studies precluded a systematic analysis of how the intracellular structure is 'seen' from a diffusing protein's point of view. Notably, the scale-dependent mobility of inert molecules that diffuse through a medium provides highly valuable information about its internal structure (Mitra *et al*., 1993; Sen, 2004) and rigidity (Mason and Weitz, 1995). In polymer solutions, both properties are linked to the behavior of the polymer chain. Corresponding measurements have been applied to assess the internal geometry and complex shear moduli of very different samples such as rocks, soils and polymer solutions (Latour *et al*., 1994; Mitra *et al*., 1993; Loskutov and Sevriugin, 2013; Ernst *et al*., 2012; Mason and Weitz, 1995; Ernst *et al*., 2014). In our previous work we introduced an approach to retrieve the intracellular structure from the scale-dependent mobility of green fluorescent protein monomers ($GFP_1$), trimers ($GFP_3$) and pentamers ($GFP_5$) in a human cell line (Baum *et al*., 2014). We measured protein transport simultaneously between hundreds of positions by multi-scale fluorescence cross-correlation spectroscopy (msFCCS) with a line-scanning confocal microscope system (Heuvelman *et al*., 2009; Baum *et al*., 2014), and evaluated the protein's mean-squared displacement (MSD) versus time. From these experiments we derived the time-dependence of the diffusion coefficient and found that both the cytoplasm and the nucleoplasm resemble a porous medium or random obstacle network for diffusing proteins.

Since the MSD is directly related to the viscoelastic properties of the medium in which Brownian motion takes place, MSD measurements can also be used to describe the cellular interior as a homogeneous linear viscoelastic medium (Mason and Weitz, 1995; Guigas *et al*., 2007). In this representation, the scale-dependence is introduced by a viscoelastic response of the medium rather than by structural constraints. Since the mobility of particles in a linear viscoelastic medium reflects the complex shear modulus (Mason and Weitz, 1995), the time-dependent diffusion coefficient can be obtained from the viscoelastic moduli of the medium and vice versa. Representing the nuclear interior as a viscoelastic medium provides a well-established set of physical parameters that integrate its liquid- and solid-like properties. These can subsequently be used in quantitative descriptions of nuclear processes that, for example, involve the rates of reactions and target search mechanisms of enzymes. Here, we set out to derive the viscoelastic properties of the nuclear interior from msFCCS experiments with $GFP_1$, $GFP_3$, $GFP_5$ and the chromatin-interacting chromodomain (CD) of heterochromatin protein 1 (HP1). We delineate the regime in which the nucleus appears as a viscous fluid on small and large scales, and where it exhibits viscoelastic properties on intermediate scales. Furthermore, we show how one can apply our approach to calculate how particles 'memorize' the history of their random walk for different times depending on their size, leading to a tracer size-dependent regulation of target search.

## 2. Methodology

Complex shear moduli revealed by different tracer molecules were calculated based on Laplace-transformed MSD measurements according to a generalized fluctuation-dissipation theorem (Mason and Weitz, 1995). Experimental MSDs were fitted with the model function given by equation (6) and Laplace-transformed with Mathematica 9.0 (Wolfram Research). A data point $P_i = (t_i, MSD_i)$ of the MSD in the time domain was transformed to Laplace space as follows: (i) The model function was fitted to a given data point by adjusting the microscopic diffusion coefficient $D_0$ while keeping the previously determined retardation $R = D_0/D_\infty$ as well as the characteristic parameter $\lambda$ of the porous medium model constant (Baum *et al.*, 2014). (ii) The Laplace transform of the adapted model function that includes the data point $P_i$ was calculated. (iii) The Laplace-transformed function at the frequency $\omega_i = 1/t_i$ was evaluated to obtain the data point $\tilde{P}_i = (\omega_i, MSD'(\omega_i))$ in the frequency domain.

Autocorrelation functions in Fig. 1c were plotted for a 3D Gaussian with lateral and axial widths of 0.25 µm and 1.25 µm, respectively, and (i) for diffusion in a fractal according to equation (1) with $\Gamma = 4\,\mu m^2 s^{-0.8}$ and $\alpha = 0.8$ (yellow), (ii) for diffusion in a porous medium according to equation (3) with $D_0 = 40\,\mu m^2 s^{-1}$, $D_\infty = 10\,\mu m^2 s^{-1}$ and $\lambda = 0.24\,\mu m$ (blue), or (iii) for diffusion in a standard linear fluid according to equation (12) with $D_\infty = \dfrac{k_B T}{6\pi \eta_2 r} = 9\,\mu m^2 s^{-1}$, $\dfrac{\eta_2}{\eta_1} = 2$ and $\dfrac{\mu}{\eta_2} = 1000\,s^{-1}$ (red).

The return preference given in Fig. 7c was calculated as the ratio between the diffusion propagator for the time-dependent diffusion coefficient $D_{porous}(t)$ defined in equation (3) and the diffusion propagator for a constant diffusion coefficient $D_0$ evaluated at distance $x = 0$. Thus, the return preference is given by $\left(\dfrac{D_0}{D_{porous}(t)}\right)^{1.5} = \left(\dfrac{D_\infty}{D_0} + \left(1 - \dfrac{D_\infty}{D_0}\right)\exp\left(-\dfrac{4\sqrt{D_0 t}}{\sqrt{\pi}\lambda}\right)\right)^{-1.5}$.

## 3. Results

### 3.1. Models for anomalous diffusion in the cell nucleus

Diffusion processes in living cells deviate from normal diffusion (Hofling and Franosch, 2013). Several models have been proposed to describe such transport processes in living cells (Fig. 1a, b), which explain deviation from free diffusion either by structural constraints of the accessible space or by a viscoelastic response of the medium.

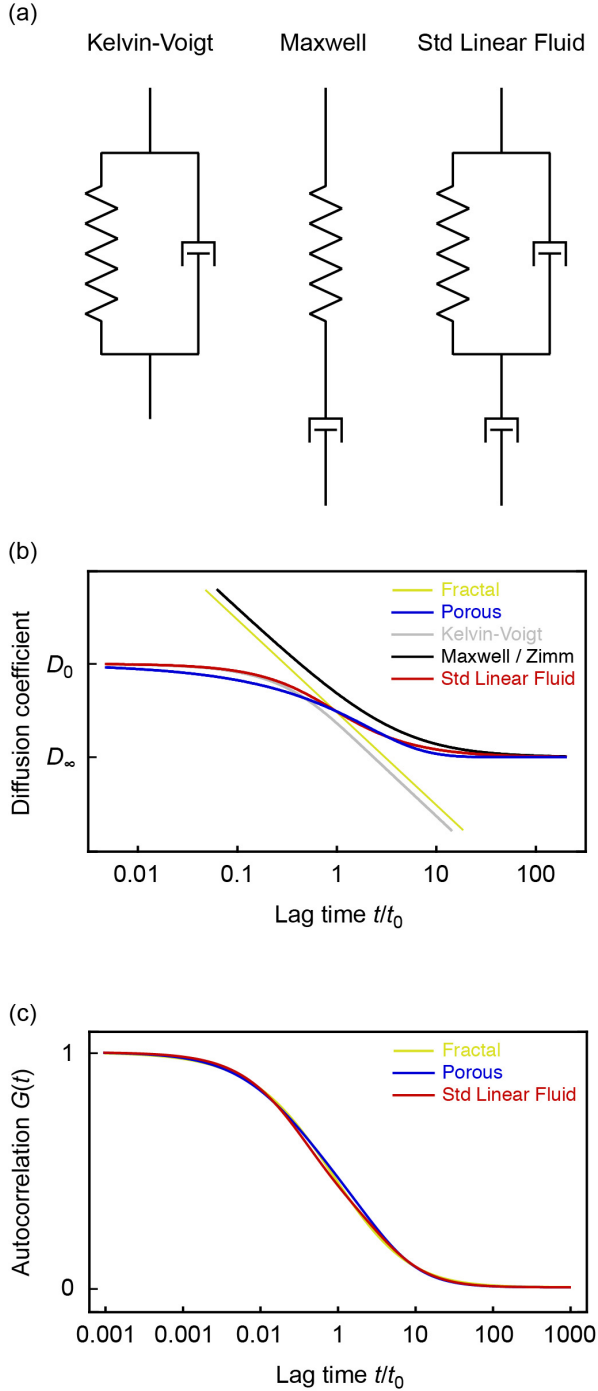

**Figure 1. Description of the nuclear interior as a viscoelastic medium.** (a) Different simple viscoelastic models that consist of elastic springs and dampers connected in parallel or in series. Left: Kelvin-Voigt model with spring and damper in parallel, which is often used to model the creep behavior of polymers at constant stress. Middle: Maxwell model with damper and spring connected in series. This model is often used to describe stress relaxation of polymers at constant strain. Right: standard linear fluid or Jeffrey fluid model that combines a Kelvin-Voigt model in series with a damper. (b) Normalized $D(t)$ profiles for different theoretical descriptions of anomalous diffusion in complex media. In contrast to normal diffusion with an invariant diffusion coefficient, the scale-dependent diffusion coefficient $D(t)$ decreases on larger time scales. It is directly related to the linear viscoelastic moduli of the medium. Structural models for a fractal environment or a porous medium are compared to viscoelastic models. For diffusion in fractals, $D(t)$ follows a straight line in the double logarithmic representation used here (yellow). For diffusion in random porous media, $D(t)$ exhibits sigmoidal shape (blue), indicating viscous behavior on small and large scales and viscoelastic behavior on intermediate scales. Simple viscoelastic models that represent the medium as a combination of springs and dampers yield viscoelastic behavior on intermediate scales but differ in their asymptotic behavior: Maxwell fluids are elastic on short scales and viscous on large scales (black), Kelvin-Voigt fluids are viscous on short scales and elastic on large scales (gray), standard linear fluids (also called Jeffrey fluids) are viscous on short and large scales (red). The Zimm model for polymer solutions is a variant of the Maxwell fluid and exhibits the same asymptotic behavior. Notably, $D(t)$ in a standard linear fluid (red) is similar to $D(t)$ in a porous medium (blue) that does not explicitly involve an elastic contribution. (c) Autocorrelation functions of particle number fluctuations in a Gaussian-shaped probe volume for different diffusion models. The predicted fluctuations are very similar and it is difficult to distinguish between different models from this type of data.

The simplest model is anomalous diffusion with a power law describing the time-dependent diffusion coefficient (Wachsmuth *et al.*, 2000; Saxton, 1994; Bunde and Havlin, 1995):

$$D_{\text{anom}}(t) = \Gamma t^{\alpha-1}. \tag{1}$$

Here, $\Gamma$ is called transport coefficient and $\alpha$ is the (constant) anomaly parameter. Accordingly, the mean squared displacement (MSD) reads

$$MSD_{anom}(t) = 6D_{anom}(t)t = 6\Gamma t^\alpha. \tag{2}$$

Since equations (1) and (2) describe diffusion on a fractal structure (Bunde and Havlin, 1995) with a fractal dimension related to the anomaly parameter $\alpha$, it was proposed that the nuclear interior might resemble a fractal (Bancaud *et al.*, 2009; Bancaud *et al.*, 2012; McNally and Mazza, 2010). However, in these experiments the anomaly parameter was mostly assessed from determining the slope of the temporally decaying autocorrelation function at a single length scale. Since different diffusion propagators can yield very similar autocorrelation functions, it is difficult to exclude other topologies or diffusion laws with this approach (Fig. 1c). An alternative is the model of diffusion in random porous media, which was developed to explain the time-dependent diffusion coefficients measured in porous materials by PFG-NMR (Loskutov and Sevriugin, 2013; Novikov *et al.*, 2011). The time-dependent diffusion coefficient in rocks, soils and biological tissues did not follow a power law but exhibited an exponential decay when plotted versus $\sqrt{t}$, which corresponds to a sigmoidal shape in double-logarithmic representation (Loskutov and Sevriugin, 2013; Latour *et al.*, 1994) (Fig. 1b). This behavior was described by the following equation

$$D_{porous}(t) = D_\infty + (D_0 - D_\infty)\exp\left(-\frac{4\sqrt{D_0 t}}{\sqrt{\pi}\lambda}\right). \tag{3}$$

Here, $D_0$ and $D_\infty$ are the diffusion coefficients on microscopic and macroscopic scales, respectively, which define the retardation $R = D_0/D_\infty$. The parameter $\lambda$ is the characteristic crossover length between both normal diffusion regimes. In the limit of short times, equation (3) converges to equation (4) (Mitra *et al.*, 1993)

$$D_{porous}^{short}(t) = D_0\left(1 - \frac{4\sqrt{D_0 t}}{9\sqrt{\pi}}\frac{S}{V_p} + O(D_0 t)\right), \tag{4}$$

with the surface-to-volume ratio

$$\frac{S}{V_p} = \frac{9}{\lambda}\left(1 - \frac{1}{R}\right). \tag{5}$$

Via equation (5) the surface-to-volume ratio of the medium is directly related to the characteristic crossover length introduced above in equation (3). Both parameters are instructive quantities that characterize the properties of the porous medium. According to equation (3), the MSD of the particles reads

$$MSD_{porous}(t) = 6D_{porous}(t)t = 6D_\infty t + 6(D_0 - D_\infty)\exp\left(-\frac{4\sqrt{D_0 t}}{\sqrt{\pi}\lambda}\right)t. \tag{6}$$

Scale-dependent diffusion is also observed in viscoelastic media that are described as a composition of springs and dampers. In this representation, the scale-dependence is introduced by an elastic response of the medium rather than by structural constraints. The following relation connects the MSD of the particles with the complex shear modulus

$$G(\omega) = G'(\omega) + iG''(\omega) = \frac{k_B T}{i\pi\omega r\, MSD'(i\omega)}. \tag{7}$$

Here, $G(\omega)$ is the complex shear modulus with $G'$ and $G''$ representing the elastic storage modulus and the viscous loss modulus, respectively, $k_B T$ is the thermal energy, $r$ is the hydrodynamic radius of the diffusing particle, and $MSD'$ is the Laplace-transformed MSD. In equation (7) the inertia of the particle is neglected and the microscopic memory function imposed by the medium is assumed to be proportional to its bulk viscosity (Mason and Weitz, 1995). Based on equation (7) the time-dependent diffusion coefficient equals

$$D(t) = \frac{k_B T}{6\pi r t} L^{-1}\left\{\frac{1}{s\, G(s)}\right\}, \tag{8}$$

where $L^{-1}\{\cdot\}$ denotes the inverse Laplace transform, and $s = i\omega$ is the Laplace variable. The complex shear modulus and the complex viscosity are connected via (Mason and Weitz, 1995)

$$\eta(\omega) = \eta'(\omega) + i\eta''(\omega) = \frac{G(\omega)}{i\omega} = \frac{G''(\omega)}{\omega} - i\frac{G'(\omega)}{\omega}. \tag{9}$$

Here, $\eta'(\omega)$ is the dynamic viscosity and $\eta''(\omega)$ corresponds to the out-of-phase viscosity that reflects the elastic response of the medium. Based on the well-known complex shear moduli for Maxwell fluids, Kelvin-Voigt fluids, standard linear fluids (SLF, also known as Jeffrey fluids) and standard linear solids (SLS) (Mainardi and Spada, 2011; Banks *et al.*, 2011; Raikher *et al.*, 2013), the following time-dependent diffusion coefficients are obtained:

$$D_{\text{Maxwell}}(t) = \frac{k_B T}{6\pi r t} L^{-1}\left\{\frac{\mu^2 - (\eta s)^2}{s(\mu^2 \eta s - \mu(\eta s)^2)}\right\} = \frac{k_B T}{6\pi r}\left[\frac{1}{\eta} + \frac{1}{\mu t}\right], \qquad (10)$$

$$D_{\text{Kelvin-Voigt}}(t) = \frac{k_B T}{6\pi r t} L^{-1}\left\{\frac{1}{s(\mu + \eta s)}\right\} = \frac{k_B T}{6\pi r \mu t}\left[1 - \exp\left(-\frac{\mu}{\eta}t\right)\right], \qquad (11)$$

$$D_{\text{SLF}}(t) = \frac{k_B T}{6\pi r t} L^{-1}\left\{\frac{\mu + (\eta_1 + \eta_2)s}{s(\eta_2 \mu s + \eta_1 \eta_2 s^2)}\right\} = \frac{k_B T}{6\pi r}\left[\frac{1}{\eta_2} + \frac{1}{\mu t}\left(1 - \exp\left(-\frac{\mu}{\eta_1}t\right)\right)\right], \qquad (12)$$

$$D_{\text{SLS}}(t) = \frac{k_B T}{6\pi r t} L^{-1}\left\{\frac{1 - (\eta \mu_1^{-1} s)^2}{s(\mu_2 + \eta s - (\mu_1 + \mu_2)(\eta \mu_1^{-1} s)^2)}\right\} = \frac{k_B T}{6\pi r \mu_2 t}\left(1 - \frac{\mu_1}{\mu_1 + \mu_2}\exp\left(-\frac{\mu_1 \mu_2}{\eta(\mu_1 + \mu_2)}t\right)\right). \qquad (13)$$

In equations (10) and (11), $\mu$ refers to the shear modulus of the springs and $\eta$ to the viscosity of the dampers. In equation (12), $\eta_1$ refers to the damper within the Kelvin-Voigt element, and $\eta_2$ to the damper in series with it. In equation (13), $\mu_1$ refers to the spring in the Maxwell element, and $\mu_2$ to the spring in parallel to it. The time-dependent diffusion coefficients for different models are plotted in Fig. 1b. All models predict viscoelastic behavior on intermediate scales and either purely viscous or purely elastic behavior on very small or large scales. Notably, both the standard linear fluid model in equation (12) and the porous medium model in equation (3) yield viscous behavior on small and large time scales. Thus, the standard linear fluid is the simplest model composed of springs and dampers that qualitatively describes porous media.

*3.2. Determination of the scale-dependent mobility and complex shear modulus from msFCCS data*
In FCS experiments, the fluctuating fluorescence signal, which is emitted by the labeled particles that move through the microscope's detection volume, is recorded over time. The temporal autocorrelation function of the signal contains information about the diffusion coefficient and the diffusion anomaly on the length scale of the diffraction-limited detection volume. Both parameters are obtained by fitting a model function to the autocorrelation curve that depends on the diffusion propagator and the geometry of the detection volume. Assuming the power law in equation (1), the width of the autocorrelation function determines the diffusion coefficient and the slope determines the anomaly parameter (Fig. 1c, 2a).

(a) Conventional FCS

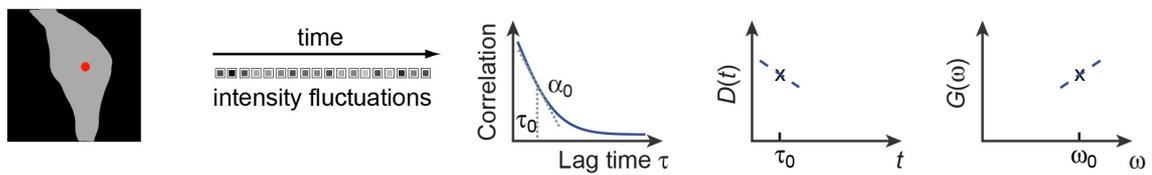

(b) Multi-scale FCCS

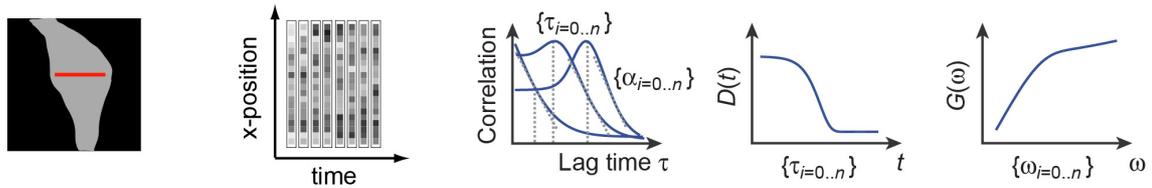

**Figure 2. Mobility measurements by fluorescence correlation spectroscopy (FCS) at a single point in comparison to measurements at multiple scales by multi-scale FCCS (msFCCS).** (**a**) In conventional FCS, the diffusion anomaly is determined from the slope of the autocorrelation curve that measures the dwell time of a particle within the diffraction-limited detection volume of a confocal microscope. Information about larger length scales is not obtained. The scale-dependent diffusion coefficient and the complex shear modulus can be extrapolated if a specific model is assumed. (**b**) The msFCCS method extends the concept of conventional FCS to multiple scales. Besides the diffusion anomaly that can be obtained on every length scale in the same manner as in conventional FCS, the translocation time is directly determined for every translocation distance along the line of the measurement. Thus, the scale-dependence of the diffusion coefficient is sampled without assumptions about the diffusion model. Accordingly, the frequency-dependent complex shear modulus is obtained on multiple scales.

Thus, both the time-dependent diffusion coefficient and the frequency-dependent shear modulus, which are connected by equation (8), can only be obtained if a specific model is assumed.

For measuring the scale-dependent mobility of particles in living cells by msFCCS, intensity fluctuations are collected from different positions arranged along a line-shaped excitation and detection volume (Fig. 2b) (Baum *et al*., 2014). By conducting a spatial cross-correlation analysis this yields apparent diffusion coefficients of tracer molecules for different separation distances. Since the environment affects the scale-dependence of the diffusion coefficient in a characteristic manner, information about the properties of the medium traversed by the tracer particles is obtained (Fig. 1b). Whereas normal diffusion is observed in viscous liquids, the diffusion coefficient in fractal, porous or viscoelastic materials decreases on larger scales in a characteristic manner. Thus, msFCCS can distinguish different diffusion models that yield similar autocorrelation functions and thus cannot be distinguished by conventional FCS (Fig. 1c). The scale-dependent diffusion coefficient can readily be converted into the complex shear modulus based on a generalized fluctuation-dissipation theorem (Mason and Weitz, 1995) as described above. Thus, a continuum representation of heterogeneous media whose internal structure reduces the mobility of tracer particles is obtained in a straightforward manner.

*3.3. Complex shear modulus in porous media*

The scale-dependent mobility of particles in viscoelastic media is intimately related to the linear viscoelastic moduli in the respective material (Mason and Weitz, 1995). To calculate the complex shear modulus based on the experimentally determined MSD of differently sized GFP multimers we used equation (7). Based on the expression in equation (6), the Laplace-transformed MSD reads

$$MSD'(i\omega) = -\frac{6(D_0 - D_\infty)}{\omega^2}\left(1 - \frac{i\omega_0}{\omega} + \left(\frac{3i}{2} + \frac{\omega_0}{\omega}\right)\sqrt{\frac{i\pi\omega_0}{\omega}} w\left(\sqrt{\frac{i\omega_0}{\omega}}\right)\right) - \frac{6D_\infty}{\omega^2} . \qquad (14)$$

Here, $w(x)$ is the Faddeeva function, and $\omega_0 = \frac{4D_0}{\pi\lambda^2}$ is the inverse crossover time. According to equations (7) and (14), the complex shear modulus is given by

$$G(\omega) = \frac{i\omega k_B T}{6\pi r\left\{(D_0 - D_\infty)\left[1 - \frac{i\omega_0}{\omega} + \left(\frac{3i}{2} + \frac{\omega_0}{\omega}\right)\sqrt{\frac{i\pi\omega_0}{\omega}} w\left(\sqrt{\frac{i\omega_0}{\omega}}\right)\right] + D_\infty\right\}} . \qquad (15)$$

For a scale-invariant diffusion coefficient $D_0 = D_\infty$, the expression for a viscous liquid is obtained:

$$G(\omega) = \frac{i\omega k_B T}{6\pi r D_0} . \qquad (16)$$

This corresponds to the Stokes-Einstein relation since the imaginary part of the complex shear modulus is the product of dynamic viscosity and frequency. To obtain expressions for the elastic storage modulus and the viscous loss modulus, equation (15) is separated into its real and imaginary parts. This yields the following expressions:

$$G'(\omega) = \frac{k_B T}{6\pi r (D_0 - D_\infty) X(\omega, \omega_0, R)} \left[ \sqrt{\frac{\pi \omega_0}{2\omega}} \left( \left( \frac{3\omega}{2} + \omega_0 \right) V(\omega, \omega_0) - \left( \frac{3\omega}{2} - \omega_0 \right) L(\omega, \omega_0) \right) - \omega_0 \right], \quad (17)$$

$$G''(\omega) = \frac{k_B T}{6\pi r (D_0 - D_\infty) X(\omega, \omega_0, R)} \left[ \frac{R\omega}{R-1} - \sqrt{\frac{\pi \omega_0}{2\omega}} \left( \left( \frac{3\omega}{2} - \omega_0 \right) V(\omega, \omega_0) + \left( \frac{3\omega}{2} + \omega_0 \right) L(\omega, \omega_0) \right) \right]. \quad (18)$$

Here, $V(\omega, \omega_0)$ and $L(\omega, \omega_0)$ are the real and imaginary Voigt functions, $R = D_0/D_\infty$ is the retardation, and $X(\omega, \omega_0, R)$ is the rescaled squared Laplace-transformed MSD (for details see appendix). For small and large frequencies, equation (15) can be approximated by the following expressions

$$G_0(\omega) = \frac{k_B T}{6\pi r D_\infty} \left[ \frac{2\sqrt{\pi}(R-1)}{\omega_0^3} \omega^4 + i\omega \right], \quad (19)$$

$$G_\infty(\omega) = \frac{k_B T}{6\pi r D_0} \left[ \frac{3(R-1)}{2R} \sqrt{\frac{\pi \omega_0 \omega}{2}} + i\omega \right]. \quad (20)$$

Accordingly, the viscoelastic phase angles for large and small frequencies are given by

$$\delta_0(\omega) = \lim_{\omega \to 0} \left( \arctan \left( \frac{1}{2\sqrt{\pi}(R-1)} \left( \frac{\omega_0}{\omega} \right)^3 \right) \right) = \frac{\pi}{2}, \quad (21)$$

$$\delta_\infty(\omega) = \lim_{\omega \to \infty} \left( \arctan \left( \frac{2R}{3(R-1)} \sqrt{\frac{2\omega}{\pi \omega_0}} \right) \right) = \frac{\pi}{2}. \quad (22)$$

This shows that the nucleus resembles a viscous fluid on small and large time scales, whereas it exhibits viscoelastic behavior on intermediate scales. The real and imaginary parts of the complex shear modulus, the complex viscosity and the viscoelastic phase angle are plotted in Fig. 3.

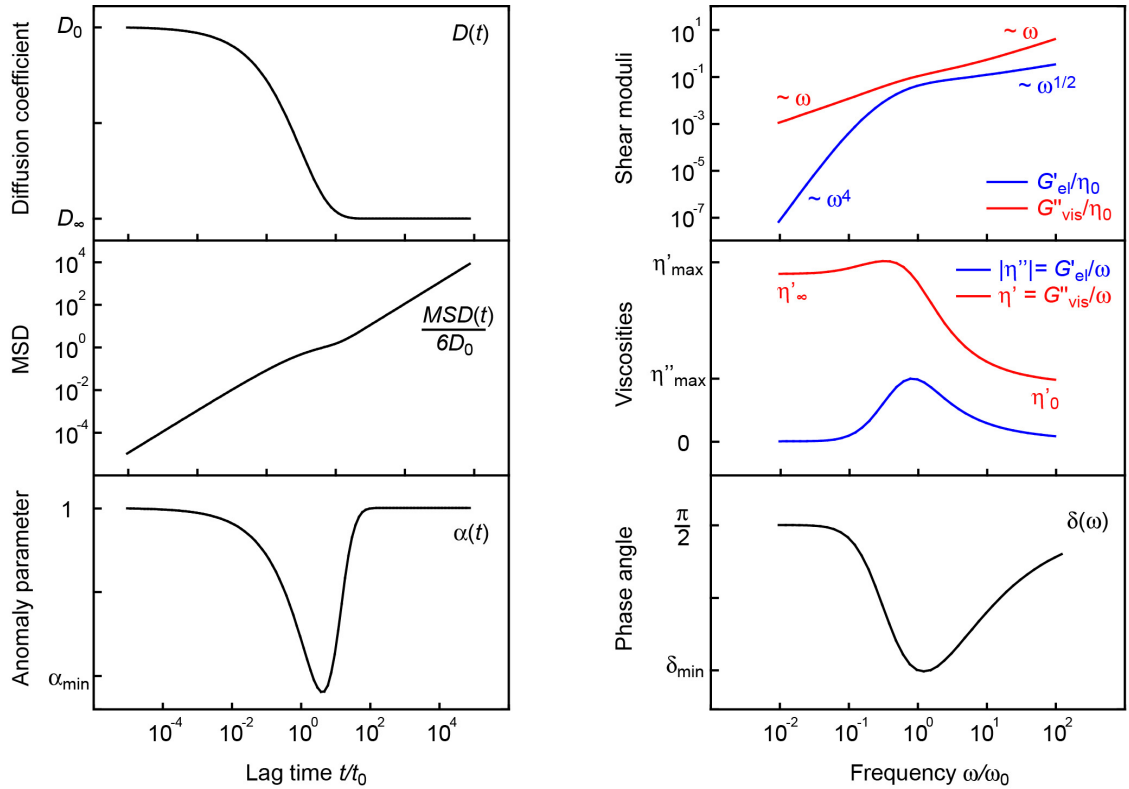

**Figure 3. From time-dependent particle mobility to frequency-dependent viscoelasticity of materials.** Based on the time-dependent diffusion coefficient $D(t)$ or the $MSD(t)$ dependence the complex shear modulus $G(\omega)$ or the complex viscosity $\eta(\omega)$ that characterize the viscoelasticity of the material can be derived. Diffusion is normal on small and large scales (anomaly parameter $\alpha = 1$) and the medium is viscous. On intermediate scales, the medium exhibits viscoelastic behavior and anomalous diffusion ($\alpha < 1$). The viscoelastic regime is characterized by a large storage modulus and a small viscoelastic phase angle $\delta(\omega)$. By measuring diffusion of particles on multiple scales, the viscoelastic behavior of a medium can be characterized.

As expected, the maximum elasticity is observed on time scales that are similar to the crossover time. For retardation $R = 1$, the equations converge to those for a normal diffusion process in a viscous liquid with vanishing elasticity and scale-invariant diffusion coefficient (Fig. 4). Interestingly, we find that the complex shear modulus scales differently than predicted by most simple models for polymer solutions and polymer melts (Zimm, 1956; Doi and Edwards, 1986). Whereas the model for porous media above predicts normal diffusion on short time scales (equations (20) and (22)), the Zimm model (Fig. 1b) for dilute polymer solutions predicts viscoelastic behavior in this regime (Guigas *et al.*, 2007; Doi and Edwards, 1986). Intuitively, normal diffusion should be obtained on length scales that are small compared to the average pore size in the polymer network. Such behavior has also been observed in numerical simulations of Brownian motion in polymers (Fritsch and Langowski, 2011). Thus, the cell nucleus is better represented by a standard linear fluid also known as Jeffrey fluid (Fig. 1a, b) that was previously found to describe the diffusion of micrometer-sized particles in highly concentrated solutions of micelles (Raikher et al., 2013).

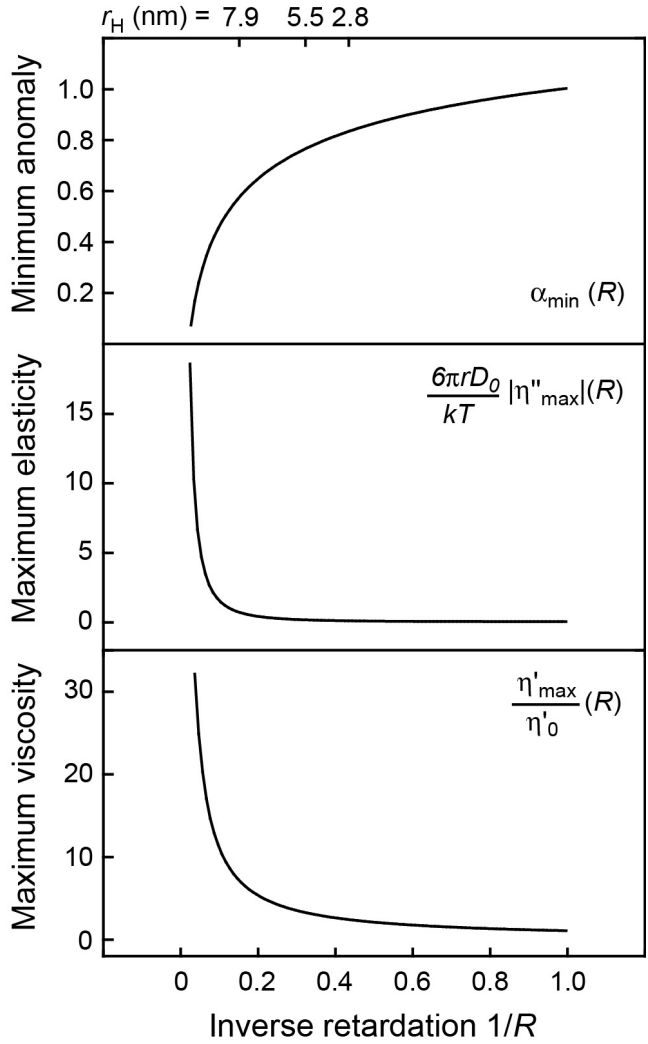

**Figure 4. Anomalous diffusion behavior and material viscoelasticity as a function of the mobility retardation.** The viscoelastic properties of a porous medium are related to the retardation $R = D_0 / D_\infty$ that corresponds to the ratio of the diffusion coefficients on small and large scales. A ratio of $R = 1$ corresponds to a purely viscous medium, in which diffusion is normal. For decreasing values of $1/R$ the elastic contribution increases and diffusion becomes more anomalous. At the top, $1/R$ values measured for $GFP_5$, $GFP_3$ and $GFP_1$ were assigned to the respective hydrodynamic radii $r_H$ of probe particles.

*3.4. Viscoelasticity of the nucleoplasm from the perspective of GFP multimers*

To study the viscoelastic properties of the cell nucleus we calculated the complex shear modulus from the MSD of differently sized GFP-tagged proteins reported previously (Baum *et al.*, 2014). Theoretical fit functions were obtained as described in the previous section, and experimental data were numerically transformed as described in the Methods section. As expected, the complex shear moduli obtained with inert $GFP_1$, $GFP_3$ and $GFP_5$ in unperturbed U2OS cells or in cells treated with trichostatin A (TSA) or chloroquine (CQ) followed the model functions given in equations (17) and (18). In contrast, the chromatin-interacting chromodomain (CD) of heterochromatin protein 1 (HP1) showed deviating behavior. The real and imaginary parts of the complex viscosity calculated according to equation (9) are shown in Fig. 5 and Fig. 6.

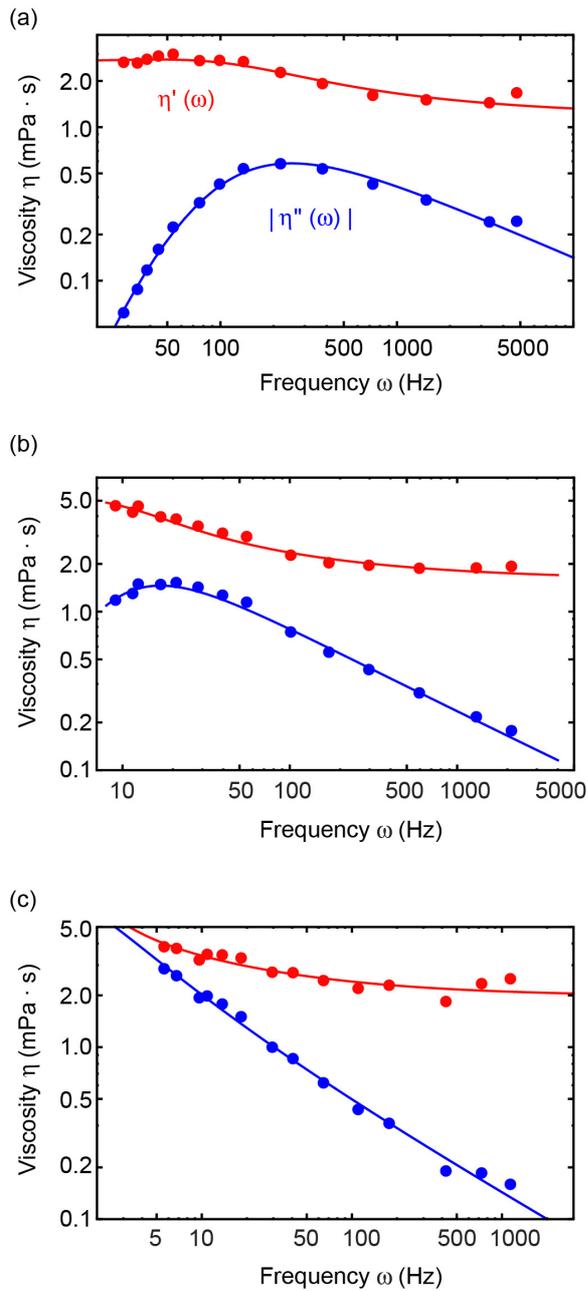
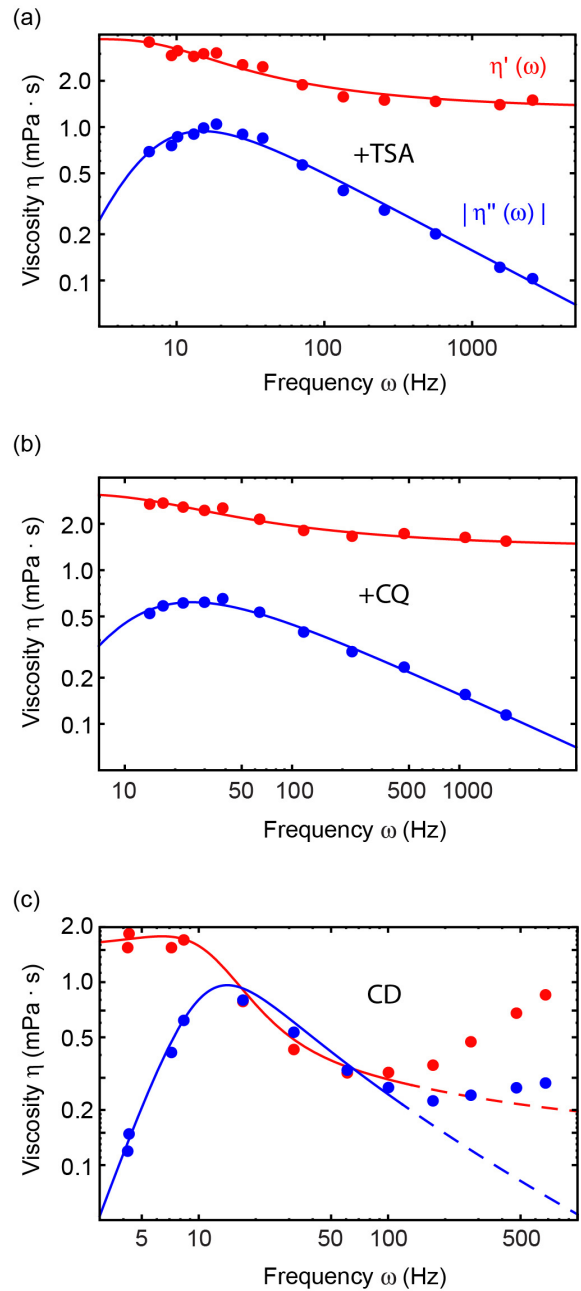

**Figure 5. Complex viscosity for GFP monomer and multimers in unperturbed cells.** The real part (red) corresponds to the dynamic viscosity and the imaginary part (blue) represents the elastic response of the nucleus. The dynamic viscosity smoothly changes between its plateau values for small and large frequencies. The elastic contribution exhibits a peak that shifts to smaller frequencies for larger particles, and vanishes for small and large frequencies. (**a**) $GFP_1$. (**b**) $GFP_3$. (**c**) $GFP_5$.

**Figure 6. Complex viscosity for $GFP_3$ in perturbed cells or in the presence of chromatin-interaction.** The complex viscosity from the perspective of $GFP_3$ was determined from measurements of $GFP_3$ in U2OS cells. Cells were treated with (**a**) the histone deacetylase inhibitor TSA or (**b**) the DNA intercalator chloroquine (CQ), which both decondense chromatin. Compared to untreated cells, both the real and the imaginary part of the complex viscosity decreased. The frequency-dependence of both parts remained similar. (**c**) The complex viscosity for the chromatin-interacting chromodomain (CD) of HP1 behaved similarly to that of $GFP_1$ for small frequencies but very differently for large frequencies.

The elastic contribution peaks near the crossover frequency (Fig. 3) that shifts to lower values for larger proteins (Fig. 5). The shear moduli at the crossover frequency amount to ~147 mPa and ~25 mPa from the perspective of $GFP_1$ and $GFP_3$, respectively (Table 1). For $GFP_5$, the crossover frequency was outside the frequency range accessible by msFCCS and therefore only limiting values were determined. For large frequencies, the real part of the complex elastic shear modulus (that corresponds to the imaginary part of the complex viscosity multiplied with the frequency) followed a power law with exponents of 0.66±0.02, 0.49±0.02 and 0.31±0.04 as determined with $GFP_1$, $GFP_3$ and $GFP_5$ in unperturbed cells. These findings indicate a more liquid-like mechanical behavior of the nuclear interior as sensed by $GFP_1$ molecules on small/length scales and a more solid-like behavior as sensed by $GFP_5$ molecules according to the assignment ($\alpha = 0$, solid-like; $\alpha = 1$, fluid-like) proposed previously (Wilhelm, 2008). The tracer size-dependence of the apparent elastic response of the nucleus is an intracellular feature that is not observed at the macroscopic scale. This might be due to the differential influence of the permeability of the nucleus and accordingly the mesh size distribution of the chromatin network on the mobility of differently sized particles. Whereas random motion of particles that are larger than the typical mesh size mainly reflects the flexibility of the chromatin chain, random motion of smaller particles is both affected by the permeability and the elasticity of the network. Thus, measurements with differently sized tracers at the micro- and macroscale yield complementary information on the intracellular environment.

The scaling of the shear modulus obtained here for large frequencies with $GFP_3$ was similar to that determined previously by conventional FCS for 5 nm gold beads tagged with streptavidin that are comparable in size (Guigas *et al.*, 2007). According to the Zimm model, an exponent of 5/9 or 2/3 indicates good or theta solvent conditions, respectively (Doi and Edwards, 1986). Here, we found that the apparent solvent conditions were dependent on protein size. Furthermore, an exponent of 0.57±0.01 determined for $GFP_3$ in CQ treated cells versus 0.49±0.02 in unperturbed cells indicated a shift towards a theta solvent environment upon chromatin decondensation. The real part of the complex viscosity exhibits sigmoidal shape in double-logarithmic representation, with plateau values of 1-2 mPa·s for large frequencies (Table 1). This is roughly twice the viscosity of water, suggesting that the viscosity of the bulk solution in the nucleoplasm is only moderately increased. In contrast, values in the range of 3-16 mPa·s were obtained for small frequencies. The larger viscosity values in the limit of small frequencies reflect the hindrance of large obstacles or additional binding reactions for the chromatin-interacting CD.

*3.5. Target search processes in the viscoelastic environment of the cell nucleus*

From the perspective of diffusing GFP monomers and multimers, the cell nucleus appears as a porous medium (Fig. 7a). In the continuum representation, it shares similarities with a linear viscoelastic fluid that is composed of a damper in series with a Kelvin-Voigt element (Fig. 7b). Although both descriptions yield similar expressions for the scale-dependent mobility of an ensemble of diffusing particles, the porous medium model captures the underlying molecular details more precisely. Histograms for the mobility of smaller subsets of particles broaden on larger length and time scales (Baum *et al*., 2014), indicating that the cell nucleus is heterogeneous. Fast and slow populations of particles can easily be explained within the framework of a porous medium, since the mobility is connected to the number and type of collisions and interactions with obstacles.

Further, we found a small fraction of $GFP_5$ molecules that became trapped, which would not be predicted in a homogenous viscoelastic medium. However, both descriptions are complementary and the continuum representation might be more efficient in simulations of transport processes and chemical reactions that account for the geometrical constraints molecules experience in the nucleus.

|  | $\omega_{viscosity}$ (Hz) | $\omega_{phase}$ (Hz) | $G'(\omega_{viscosity})$ (mPa) | $\eta_0$ (mPa·s) | $\eta_\infty$ (mPa·s) |
|---|---|---|---|---|---|
| $GFP_1$ | 253 | 412 | 147 | 1.2 ± 0.1 | 2.7 ± 0.3 |
| $GFP_3$ | 17 | 30 | 25 | 1.6 ± 0.1 | 5.0 ± 0.6 |
| $GFP_5$ | > 0.1 [x] | > 0.3 [x] | < 6 [x] | 2.1 ± 0.2 [x] | < 69 [x] |
| $GFP_3$ + TSA | 15 | 26 | 14 | 1.3 ± 0.1 | 3.6 ± 1.0 |
| $GFP_3$ + CQ | 25 | 40 | 15 | 1.4 ± 0.1 | 3.1 ± 0.5 |
| $CD$-$GFP_1$ | 14 | 29 | 135 | 1.5 * | 15.6 ± 6.3 |

**Table 1.** Summary of microscopic and macroscopic viscosities $\eta_0$ and $\eta_\infty$, elastic shear moduli *G'* as well as resonance frequencies for the maximum out-of-phase viscosity $\omega_{viscosity}$ and for the minimum viscoelastic phase angle $\omega_{phase}$ determined here. Results are based on measurements of $GFP_1$, $GFP_3$ and $GFP_5$ in the nucleus of unperturbed living cells. Further, the viscoelasticity of the cell nucleus following treatment with trichostatin A (TSA) or chloroquine (CQ) was determined for $GFP_3$. In unperturbed cells, values for the chromatin-interacting chromodomain (CD) are listed.
* The microscopic diffusion coefficient was fixed for least squares fitting to that of $CD$-$GFP_1$ in the cytoplasm since these values should be similar on small time-scales.
[x] The time-dependence of the MSD of $GFP_5$ molecules could not be measured on large enough time scales for determining the macroscopic diffusion coefficient. Only a lower limit could be estimated, leading to upper or lower limits for the marked values.

Based on the diffusion propagator for porous media the influence of collisions with obstacles on target search processes can be evaluated. Due to the compaction of the random walks particles become confined and preferentially sample sites in their local vicinity of size $\lambda$ (Fig. 7c). Since this confinement effect depends on the retardation *R* that changes with particle size, smaller particles behave differently than larger particles. This is shown here by means of the survival probabilities within a corral of size $\lambda$ for $GFP_1$, $GFP_3$ and $GFP_5$. Compared to a purely viscous medium without obstacles, particles reside considerably longer within their corrals. Consistently, the probability for a particle to return to its initial position is enhanced in the nucleoplasm compared to a viscous fluid (Fig. 7c, see Methods section for mathematical description). As discussed below, this probability might be functionally relevant for processes in which enzymes carrying a short-lived modification dissociate and re-associate to a new target site.

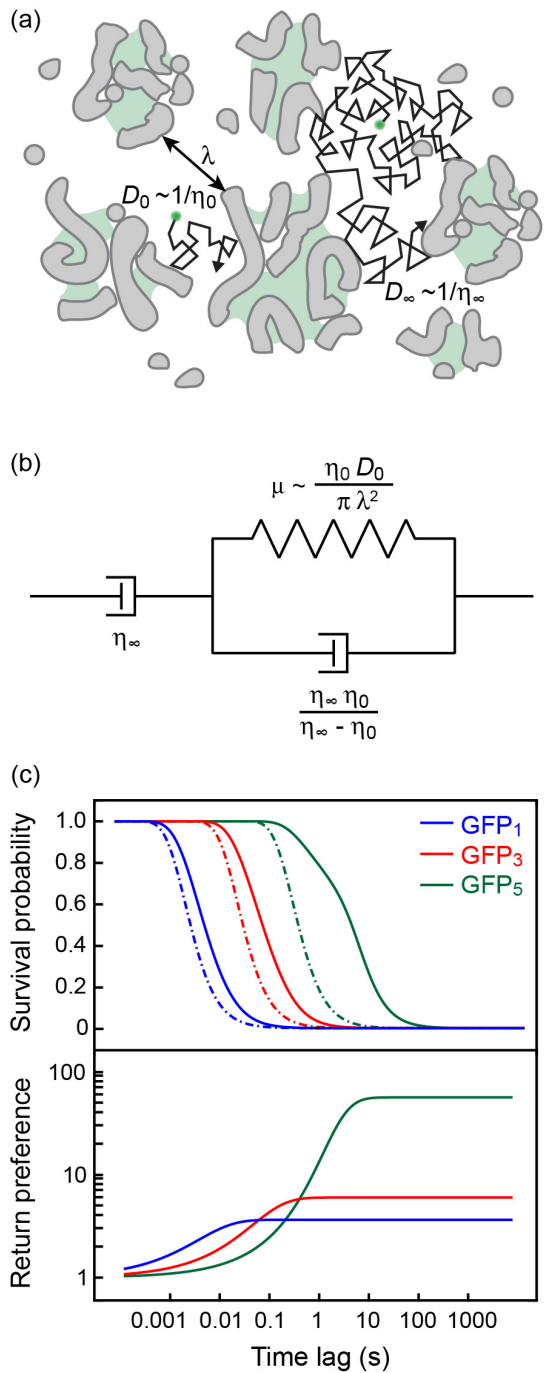

**Figure 7. The nuclear interior as a viscoelastic medium** (**a**) The cell nucleus resembles a porous medium composed of polymeric macromolecules acting as obstacles that compact the random walk of diffusing particles. On small and large time scales, the nucleoplasm appears purely viscous and particles diffuse with $D_0$ and $D_\infty$, respectively. On intermediate scales, the elastic contribution increases and the medium is viscoelastic. The location of this regime changes with particle size. (**b**) The viscoelastic model for a standard linear fluid (or Jeffrey fluid), which consists of a damper in series with a Kelvin-Voigt element, reproduces the behavior of a porous medium. The viscosity of the dampers and the shear modulus of the spring are related to the viscosity of the bulk fluid and the characteristic crossover length in the porous medium model as indicated. (**c**) Compared to a purely viscous medium (dash-dotted line), particles in a porous medium (solid line) remain longer in their local environment as inferred from the survival probabilities in a region of size $\lambda$. Further, the probability for a particle to return to its initial position is enhanced in porous media. For both descriptions of the nuclear interior the efficiency of target search processes is affected as compared to purely viscous media.

## 4. Discussion

In the present study we have developed a quantitative description of the nuclear interior as a viscoelastic medium. It describes the reduction of the particles' apparent diffusion coefficient and the compaction of random walks by a continuum representation. In the homogenous viscoelastic medium, an elastic response acts on diffusing particles and collisions with large obstacles like chromatin are modeled as an elastic force. We determined the complex shear modulus from the perspective of differently sized proteins in the nucleus based on previously reported MSD measurements by msFCCS (Baum *et al.*, 2014). The elastic contribution of the nuclear interior exhibited a maximum on characteristic time scales of 4 ms and 59 ms for $GFP_1$ and $GFP_3$, respectively, with shear moduli of 147 mPa and 25 mPa. Somewhat larger values of 570 mPa at 10 ms were previously obtained by conventional FCS experiments with gold beads in the nucleus of HeLa cells (Guigas *et al.*, 2007). The deviation might be caused by the assumption of a power law for the MSD and the complex shear modulus in the latter study, which does not fit the scaling behavior that we observed for GFP multimers by msFCCS (Baum *et al.*, 2014). Furthermore, msFCCS yielded absolute diffusion coefficients that were roughly 1.5-fold larger than the values obtained previously by Guigas et al. (Guigas *et al.*, 2007). Since the viscoelastic regime determined here comprised protein size-dependent frequency ranges, our data are incompatible with a fractal organization of the nuclear interior (Saxton, 1993).

The simplest linear viscoelastic model that reproduced the key features of our measurements is the standard linear fluid or Jeffrey fluid (Fig. 7b): It models the medium as a damper in series with a Kelvin-Voigt element (consisting of a damper and a spring in parallel). The viscoelastic behavior of the nucleus has important implications for enzymatic reactions and target search processes. Compared to purely viscous fluids, particles tend to stay in a corral of distinct size, which favors rebinding of targets in their local vicinity. This might be particularly relevant for reaction-diffusion processes, in which multiple association-dissociation events occur sequentially within a spatially confined region. One example might be transcription by RNA polymerase that occurs more efficiently when the transcription start and termination sites are located in proximity to promote multiple transcription cycles by the same RNA polymerase complex (Perkins *et al.*, 2008; Tan-Wong *et al.*, 2008; Larkin *et al.*, 2013). According to the considerations above, such cycles of dissociation and rebinding are favored in a viscoelastic medium. Furthermore, an increased return preference to previous binding sites might enhance chromatin clustering by DNA binding proteins (Brackley *et al.*, 2013) and interactions between chromatin loci mediated by diffusible binding proteins (Barbieri *et al.*, 2012). Moreover, the viscoelastic nature of the nuclear environment may promote the establishment of specific genomic interactions needed for recombination events (Lucas *et al.*, 2014).

In summary, our analysis shows that the cell nucleus resembles a viscous liquid on small and large scales. On intermediate scales it exhibits viscoelastic properties that change with particle size. The theoretical description presented here enables the construction of simplified models that reproduce the constraints imposed on the mobility of particles that diffuse within the nucleus. Such models are important for our understanding of reaction-diffusion and target search processes that occur in the nucleus.

## 5. Acknowledgements

This work was supported by the project ImmuoQuant (0316170B) of the German Federal Ministry of Education and Research (BMBF) and a DKFZ intramural grant to FE.

## 6. Appendix

To obtain the expressions for the elastic and viscous moduli, the Laplace-transformed MSD in equation (14) is decomposed into its real and imaginary part:

$$\text{Re}\{MSD'\} = -\frac{6(D_0 - D_\infty)}{\omega^2}\left(1 + \sqrt{\frac{\pi\omega_0}{2\omega}}\left[\left(\frac{\omega_0}{\omega} - \frac{3}{2}\right)V(\omega,\omega_0) - \left(\frac{\omega_0}{\omega} + \frac{3}{2}\right)L(\omega,\omega_0)\right]\right) - \frac{6D_\infty}{\omega^2} \quad (A23)$$

$$\text{Im}\{MSD'\} = -\frac{6(D_0 - D_\infty)}{\omega^2}\left(-\frac{\omega_0}{\omega} + \sqrt{\frac{\pi\omega_0}{2\omega}}\left[\left(\frac{\omega_0}{\omega} - \frac{3}{2}\right)L(\omega,\omega_0) + \left(\frac{\omega_0}{\omega} + \frac{3}{2}\right)V(\omega,\omega_0)\right]\right) \quad (A24)$$

Here, $V$ and $L$ are abbreviations for the real and imaginary Voigt function, respectively, which equal the real and imaginary part of the Faddeeva function:

$$V = V(\omega,\omega_0) = \frac{1}{\sqrt{\pi}}\int_0^\infty \exp\left(-\frac{t^2}{4}\right)\exp\left(-\sqrt{\frac{\omega_0}{2\omega}}t\right)\cos\left(\sqrt{\frac{\omega_0}{2\omega}}t\right)dt \quad (A25)$$

$$L = L(\omega,\omega_0) = \frac{1}{\sqrt{\pi}}\int_0^\infty \exp\left(-\frac{t^2}{4}\right)\exp\left(-\sqrt{\frac{\omega_0}{2\omega}}t\right)\sin\left(\sqrt{\frac{\omega_0}{2\omega}}t\right)dt \quad (A26)$$

Accordingly, the squared Laplace-transformed MSD reads

$$\left(MSD'\right)^2 = \operatorname{Re}\{MSD'\}^2 + \operatorname{Im}\{MSD'\}^2 = \frac{36(D_0 - D_\infty)^2}{\omega^4} X(\omega, \omega_0, R) \tag{A27}$$

with the abbreviation $X(\omega, \omega_0, R)$ introduced in equations (17) and (18)

$$\begin{aligned}X(\omega,\omega_0,R) =& \left(\frac{R}{R-1}\right)^2 - \frac{3R}{2(R-1)}(L+V)\sqrt{\frac{2\pi\omega_0}{\omega}} + \left(\frac{9\pi}{4}\left(L^2+V^2\right) + \frac{R-3}{2(R-1)}(L-V)\sqrt{\frac{2\pi\omega_0}{\omega}}\right)\frac{\omega_0}{\omega} \\ & + \left(1 - (L+V)\sqrt{\frac{2\pi\omega_0}{\omega}}\right)\left(\frac{\omega_0}{\omega}\right)^2 + \pi\left(L^2+V^2\right)\left(\frac{\omega_0}{\omega}\right)^3\end{aligned} \tag{A28}$$

Here, $R = D_0/D_\infty$ is the retardation, and $V$ and $L$ are abbreviations for the real and imaginary Voigt functions that are defined in equations (A25) and (A26).